# Об измерениях коэффициентов ударной ионизации электронов и дырок в 4H-SiC

## А. С. Кюрегян

Всероссийский Электротехнический институт им. В. И. Ленина, 111250, Москва, Россия
E-mail: semlab@yandex.ru

Проведен анализ всех опубликованных результатов измерений коэффициентов ударной ионизации электронов $\alpha_n$ и дырок $\alpha_p$ в 4H-SiC при 300К. Показано, что наиболее правдоподобные аппроксимации зависимостей $\alpha_{n,p}$ от напряженности электрического поля $E$ имеют обычный вид $\alpha_{n,p} = a_{n,p} \exp(-E_{n,p}/E)$ при значениях подгоночных параметров $a_n = 38.6 \cdot 10^6$ см$^{-1}$, $E_n = 25.6$ МВ/см, $a_p = 5.31 \cdot 10^6$ см$^{-1}$, $E_p = 13.1$ МВ/см. Эти зависимости $\alpha_{n,p}(E)$ использованы для расчета максимальной напряженности поля $E_b$ и толщины области пространственного заряда $w_b$ при напряжении пробоя $U_b$. Получен ряд новых формул для вычисления $\alpha_{n,p}(E)$ из результатов измерения коэффициентов лавинного умножению и факторов избыточного шума при одностороннем освещении фотодиодов со ступенчатым легированием.

Измерению зависимостей коэффициентов ударной ионизации электронов $\alpha_n$ и дырок $\alpha_p$ в 4H-SiC от напряженности электрического поля $E$ посвящено значительное количество работ [1-12]. Большинство результатов получено при комнатной температуре и ориентации поля вдоль оси $C$ кристаллов. В настоящей работе мы ограничимся анализом именно этих результатов. Зависимости $\alpha_{n,p}(E)$ аппроксимировались авторами функциями

$$\alpha_i = a_i \frac{E}{E_i} \exp\left[-\frac{E_i}{E(b_i + E/E_i)}\right], \tag{1}$$

$$\alpha_i = a_i \exp\left[-\left(\frac{E_i}{E}\right)^{b_i}\right], \tag{2}$$

где $a_i, b_i, E_i$ - подгоночные параметры, значения которых приведены в таблице, $i = n, p$. Результаты расчетов $\alpha_{n,p}(E)$ по этим формулам изображены на Рис. 1. Как видно, измерения разных авторов отличаются друг от друга в несколько раз, а при экстраполяции в область «слабых» полей, характерных для мощных приборов, различия достигают порядков величины. Две работы были посвящены проверке достоверности опубликованных ранее зависимостей $\alpha_{n,p}(E)$: в работе [13] подтверждены результаты [3], а результаты измерений [14] хорошо согласованы с данными [1,2] и противоречат данным [3] и [9].

Главная проблема измерений $\alpha_{n,p}(E)$ в SiC обусловлена тем, что совершенные слаболегированные кристаллы изготавливаются исключительно путем эпитаксиального наращивания относительно тонких пленок на сильно легированные подложки, толщина которых много больше диффузионной длины неосновных носителей заряда. Это не позволяет осуществлять «чистую» инжекцию неосновных носителей заряда в область пространственного заряда (ОПЗ) p-n-переходов с помощью фотоионизации со стороны подложки и реализовать классический метод определения $\alpha_{n,p}(E)$ [15], основанный на **раздельном** измерении коэффициентов умножения электронов $M_n$ и дырок $M_p$ в **одном и том же** диоде. Поэтому авторы большинства ра-бот были вынуждены измерять и использовать зависимости фототока $J_{ph}$ от напряжения $U$, полученные при освещении одной из поверхностей диодов излучением с разными длинами волн $\lambda = (244-375)$ nm и разными коэффициентами поглощения $\alpha_{ph} = (40-4\cdot10^4)$ cm$^{-1}$. При больших $\alpha_{ph}$ фотоионизация происходит в тонком приповерхностном слое и в ОПЗ инжектируются только электроны или только дырки. При малых $\alpha_{ph}$ фотоионизация однородна по толщине прибора и умножение получается смешанным. Использовалась также дополнительная информация (зависимость напряжения пробоя $U_b$ от концентрации легирующих примесей [10] и зависимости фактора избыточного шума $F$ от $M$ и $\lambda$ [3,4,7]), которая при правильном использовании может быть полезной для определения $\alpha_{n,p}(E)$.

| Источник | $10^{-6} a_n$, см$^{-1}$ | $E_n$, МВ/см | $b_n$ | $10^{-6} a_p$, см$^{-1}$ | $E_p$, МВ/см | $b_p$ | Формула |
|---|---|---|---|---|---|---|---|
| [1,2] | 0,632 | 6.32 | 0 | 0,70 | 4,88 | 0,227 | (1) |
| [3,4] | 1,98 | 9,46 | 1,42 | 4,38 | 11,4 | 1,06 | (2) |
| [5,6] | 2,78 | 10,5 | 1,37 | 3,51 | 10,3 | 1,09 | (2) |
| [7] | 0,0193* | 2,89* | 4,83* | 0,06 | 1,387 | 0,96 | (2) |
| | 0,0188** | 9,13** | 1,46** | | | | |
| [8] | 8190,0 | 39,4 | 1,0 | 4,48 | 12,8 | 1,0 | (2) |
| [9] | - | - | - | 3,25 | 17,9 | 1,0 | (2) |
| [10] | 176,0 | 33,0 | 1,0 | 340,0 | 25,0 | 1,0 | (2) |
| [11,12] | 3,36 | 22,6 | 1,0 | 8,5 | 15,97 | 1,0 | (2) |
| Усреднение данных работ [1-8] | **38,6±15** | **25,6±0,1** | **1,0** | **5,31±0,3** | **13,1±0.01** | **1,0** | (2) |
| | 5,96±4 | 1,43±0,6 | 1,2±0,2 | 4,11±0,7 | 1,17±0,1 | 1,04±0,03 | (2) |

Примечания:

\* - при $E < 2.5$ МВ/см

\*\* - при $E > 2.5$ МВ/см

К сожалению, методы обработки экспериментальных зависимостей $J_{ph}(U)$ описаны в [1-12] недостаточно подробно для того, чтобы можно было как следует оценить достоверность окончательных результатов. Эти методы представляются нам либо непонятными, либо неточными, либо ошибочными. Во всяком случае можно утверждать, что в большинстве[1] опубликованных работ не было представлено убедительных доказательств выполнения всей совокупности жестких требований, обеспечивающих необходимую точность результатов. Например, давно известно [16-18], что для вычисления коэффициентов умножения (особенно близких к 1) из измеренных зависимостей $J_{ph}(U)$ необходимо использовать точную и адекватную каждому конкретному эксперименту модель, описывающую $J_{ph}(U)$ без учета ударной ионизации. Между тем такая модель достаточно внятно описана только в работах [2,8], в работах [3-6] использовалась линейная экстраполяция функции $J_{ph}(U)$ из области малых напряжений, а в работах [9-12] эта проблема вообще не затронута. Кроме того, в большинстве работ использовались неудовлетворительные методы вычислении $\alpha_{n,p}$ (в работах [1-6] использовалось априорное предположение о том, что зависимости $\alpha_{n,p}(E)$ имеют вид (2), в [7] использовались ошибочные формулы (см. далее)), а в [10-12] эти методы вообще не описаны.

---

[1] Отрадным исключением является последняя опубликованная работа [8]. Однако, в ней представлены результаты измерений в относительно узком диапазоне полей (особенно это касается измерений $α_n$), которые дополняют, но не могут заменить всей совокупности результатов других авторов.

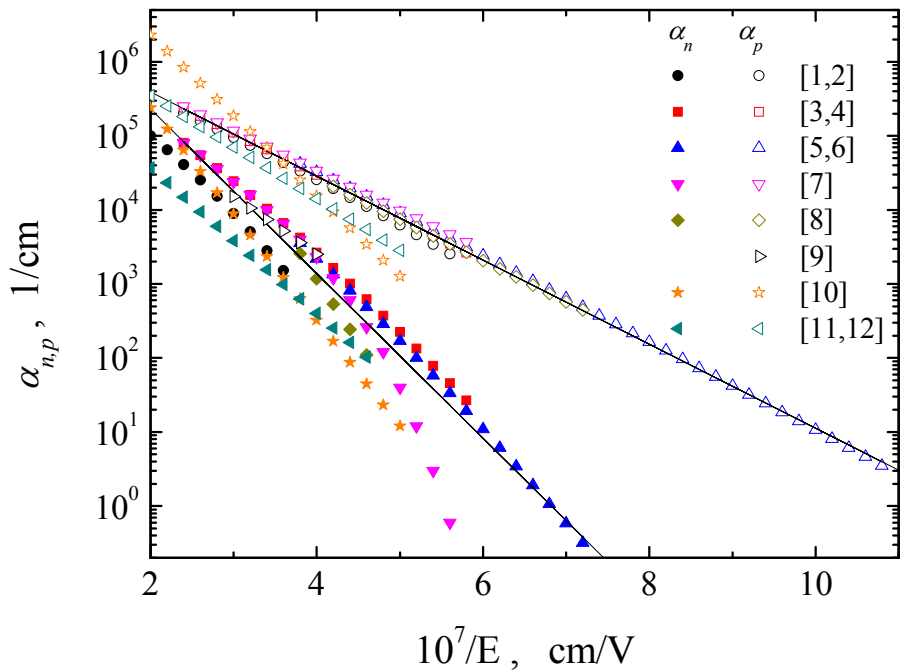

Рис. 1. Символы – значения коэффициентов ударной ионизации $\alpha_{n,p}$ в 4H-SiC, полученные по формулам (1) и (2) при $E = E_m = (10\text{ MV/cm})/(2 + 0.2m)$ и целых числах $m$, соответствующих указанным авторами [1-12] интервалам полей, в которых проводились измерения. Линии – расчет по формуле (2) при $b_i = 1$ и усредненных значениях $a_i, E_i$.

В этих условиях не представляется возможным выделить среди опубликованных зависимостей $\alpha_{n,p}(E)$ «самые надежные», которые можно было бы рекомендовать для использования. Однако по нашему мнению выход из сложившейся неопределенной ситуации существует и состоит в следующем. На окончательный результат измерений $\alpha_{n,p}(E)$ влияет более десятка различных факторов, часть из которых перечислена, например, в [19]. Упомянутые выше систематические ошибки, возникающие вследствие неполного или неточного учета этих факторов, могут иметь любой знак и изменяться в широких пределах в зависимости от деталей методики измерений, конструкции образцов и процедуры обработки данных. В этом отношении все работы отличаются друг от друга, причем почти неконтролируемым образом. Поэтому при большом количестве результатов систематические ошибки можно рассматривать как независимые случайные величины, среднее значение которых близко к нулю [20]. Такое предположение позволяет определить наиболее правдоподобные зависимости $\alpha_{n,p}(E)$ с помощью стандартных процедур усреднения.

Для этого мы использовали результаты работ [1-8], в которых получены очень близкие зависимости $\alpha_p(E)$. Такое совпадение указывает на отсутствие в этих работах грубых ошибок, поскольку вследствие сильного неравенства $\alpha_p \gg \alpha_n$ для определения $\alpha_p(E)$ достаточно использовать одну экспериментальную зависимость $M_p(U)$, исключая проблемы, связанные с наличием толстой подложки. Усреднение проводилось путем линейной регрессии на множестве точек $\{\ln[\alpha(E_m)], 1/E_m\}$, где величины $\alpha(E_m)$ вычислялись по формулам (1) или (2) с параметрами из Таблицы при значениях $E_m = (10\text{ MV/cm})/(2 + 0.2m)$. Целые числа $m$ соответствовали указанным в каждой из работ [1-8] интервалам полей, в которых проводились измерения. В этом случае получались наилучшие аппроксимации (2) при условиях $b_i = 1$. Аналогично проводилось усреднение путем нелинейной регрессии при свободных значениях $b_i$. Результаты ус-реднения и погрешности определения параметров приведены в Таблице. Как видно, учет возможного отклонения $b_i$ от 1 не имеет смысла, так как не приводит к повышению точности аппроксимации (2). Поэтому мы полагаем, что в настоящее время для зависимостей $\alpha_{n,p}(E)$ в 4H-SiC при 300 К наиболее правдоподобной является аппроксимация (2) с параметрами, которые выделены жирным шрифтом в таблице.

В заключение приведем результаты расчета напряжения $U_{jb}$, максимального поля $E_{jb}$ и толщины ОПЗ $w_{jb}$ при пробое асимметричных $p^+ - n$ ($j=d$), $n^+ - p$ ($j=a$) и $p$-$i$-$n$-диодов ($j=0$) на основе 4H-SiC с использованием новых зависимостей $\alpha_{n,p}(E)$. Основной результат заключается в том, взаимосвязь между $U_{jb}$ и $E_{jb}$ можно (см. Рис. 2), так же как в случаях Si и 6H-SiC [20], описывать соотношением

$$\frac{E_{js}}{E_{jb}} = \ln\frac{U_{jb}}{U_{js}}. \qquad (3)$$

Его погрешность не превосходит 2% в диапазоне значений $U_{jb} = 30\text{V} \div 60\text{kV}$ при значениях подгоночных параметров $E_{as} = E_{ds} = 15.9$ МВ/см, $U_{ds} = 2.15$ В, $U_{as} = 1.72$ В, $E_{0s} = 13.9$ МВ/см и $U_{as} = 1.05$ В. Используя (3) и известные формулы

$$N_j = \frac{\varepsilon}{q}\frac{E_{jb}^2}{2U_{jb}}, \qquad w_{jb} = 2\frac{U_{jb}}{E_{jb}}, \qquad (4)$$

можно вычислить концентрацию примесей $N_j$ и толщину ОПЗ $w_{jb}$ при пробое, соответствующие каждому значению $U_{jb}$. Результаты таких вычислений приведены на Рис. 3. Различия между значениями $U_{ds}$ и $U_{as}$ приводит к тому, что напряжения пробоя $p^+ - n -$ и $n^+ - p -$ переходов не совпадают при одинаковых концентрациях легирующей примеси (см. вкладку на Рис. 2). Причина этого состоит в том, что отношение $\alpha_n/\alpha_p$ зависит от напряженности поля [18].

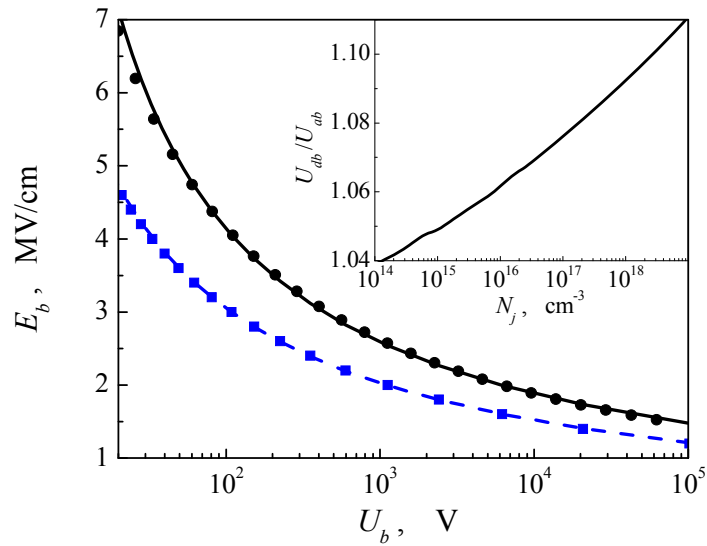

Рис. 2. Зависимость максимальной напряженности поля при пробое асимметричных $p^+ - n -$ диодов (кружки и сплошная линия) и $p$-$i$-$n$-диодов от напряжения пробоя (квадраты и штриховая линия). Символы – результат точного вычисления на основе критериев пробоя, линии – расчет по формуле (3). На вставке приведена зависимость отношения напряжений пробоя $p^+ - n -$ и $n^+ - p -$ переходов от концентрации легирующей примеси.

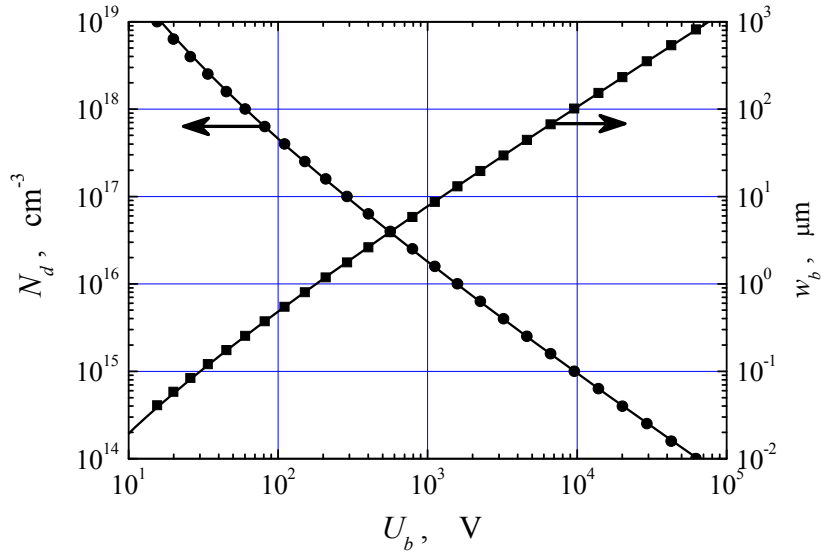

Рис. 3. Зависимость концентрации примесей и толщины ОПЗ при пробое асимметричных $p^+ - n$ – переходов от напряжения пробоя. Символы – результат точного вычисления на основе критерия пробоя, линия – расчет по формулам (3), (4).

Разумеется, эти результаты нельзя считать окончательными. Измерения $\alpha_{n,p}(E)$ в таком важном и перспективном материале, как 4H-SiC должны и будут продолжены с использованием все более совершенных методов измерения и обработки данных. Поэтому мы считаем целесообразным привести в Приложении несколько простых и, насколько нам известно, новых формул, которые позволяют вычислять зависимости $\alpha_{n,p}(E)$ в «полупроводниках на толстых подложках» без априорных предположений о характере этих зависимостей.

*Приложение.*

**Методы вычисления коэффициентов ударной ионизации.**

Известно [21], что коэффициенты ударной ионизации $\alpha_{n,p}$ связаны с фототоком $J_{ph}$ соотношением

$$(J_{ph} - J_{n0})\Phi(w) = J_{p0} + \int_0^w (J_{ph}\alpha_n + qG)\Phi(x)dx \qquad (5)$$

где $J_{n0}$ и $J_{p0}$ - токи электронов и дырок, инжектированных в ОПЗ $n^+ - n - p^+$ – структуры справа и слева соответственно, $G = gS$, $g$ - скорость фотоионизации в ОПЗ, $S$ - освещаемая площадь, $w$ - толщина ОПЗ,

$$\Phi(x) = \exp\left[\int_0^x (\alpha_n - \alpha_p)dy\right], \qquad \Phi(w) = M_n/M_p, \qquad (6)$$

$M_n = J_{ph}/J_{n0}$ при $J_{p0} = 0 = G$ и $M_p = J_{ph}/J_{p0}$ при $J_{n0} = 0 = G$. Если $n$ – слой однородно легирован донорами с концентрацией $N_d$, а его толщина $W > w$, то дифференцирование (5) и (6) по $w$ приводит к известным [22] формулам

$$\alpha_p(E_M) = \frac{M_p'}{M_n M_p}, \qquad \alpha_n(E_M) - \alpha_p(E_M) = \frac{M_n'}{M_n} - \frac{M_p'}{M_p}, \qquad (7)$$

где «штрих» означает операцию $d/dw$.

Если фотодиод изготовлен на $n^+$ – подложке, то реализовать «чистую» инжекцию дырок в ОПЗ крайне затруднительно [17]. В этом случае кроме $M_n$ можно измерять фототок $J_{ph}$ при однородной по толщине n-слоя фотоионизации. В этом случае $M_n$ и $J_{ph}$ связаны с $\alpha_{n,p}$ соотношениями

$$\frac{M'_n}{M_n} = \alpha_p(E_M)(M_n - 1) + \alpha_n(E_M), \tag{8}$$

$$\frac{dJ'_{ph}}{M_n} - (J_{ng} + J_{pg})' = qG + J_{ph}\alpha_p(E_M) + J_{ng}[\alpha_n(E_M) - \alpha_p(E_M)], \tag{9}$$

где $J_{ng}$ и $J_{pg}$ - фототоки электронов и дырок, инжектируемых в ОПЗ при однородной фотоионизации. Первое из них следует прямо из (7), а второе получается в результате дифференцирования (5) и несложных преобразований с учетом (8). Если толщина $p^+$ – слоя $l_p \ll w + L_p$, где $L_p$ - диффузионная длина дырок в $n$ – слое, то при $\alpha_p \gg \alpha_n$ (как в 4H-SiC) последним слагаемым в правой части (9) можно пренебречь, так как в этом случае $J_{ph} \gg J_{ng}$. Если еще $L_p < (W - w)$ вплоть до напряжения пробоя, то ток $J_{pg}$ не должен зависеть от напряжения и тогда можно пренебречь вторым слагаемым в левой части (9). В этом случае до начала умножения фототок $J_{ph}$ должен зависеть от $w$ по линейному закону $J_{ph} = J_{ng} + J_{pg} + qGw$, наблюдавшемуся авторами работы [2]. Тогда вместо (9) можно использовать более простую формулу

$$\alpha_p(E_M) = \frac{J'_{ph} - qGM_n}{M_n J_{ph}}, \tag{10}$$

определяя величину $qG$ по наклону зависимости $J_{ph}(w)$.

Если фотодиод изготовлен на $p^+$ – подложке, то можно измерять можно $M_p$ и $J_{ph}$ и для вычисления $\alpha_{n,p}$ использовать формулы

$$\alpha_p(E_M) = \frac{qGM'_p}{M_p J'_{ph} - M'_p J_{ph}}, \tag{11}$$

$$\alpha_n = \alpha_p - \frac{M'_p}{M_p} + \left(\frac{M'_p}{\alpha_p M_p}\right)', \tag{12}$$

которые следуют из (7), (10). Последняя из них содержит вторую производную $M''_p$, что, конечно, затрудняет обработку результатов измерений.

Следует иметь в виду, что такой способ определения $\alpha_{n,p}$ пригоден только тогда, когда можно пренебречь изменением коэффициента поглощения света в ОПЗ вследствие эффекта Франца-Келдыша. Чтобы избежать связанных с этим осложнений, можно вместе с $M_{n(p)}$ измерять факторы избыточного шума $F_{n(p)}$, которые равны

$$F_n = 2 - M_n + 2\frac{M_p^2}{M_n}\int_0^w \alpha_n \Phi^2(x) dx \tag{13}$$

при чистой инжекции электронов и

$$F_p = 2 - M_p + 2M_p \int_0^w \alpha_p \Phi^2(x) dx \tag{14}$$

при чистой инжекции дырок [23]. Из (13) и (14) нетрудно получить общее соотношение

$$M_n(F_n - 2) = M_p(F_p - 2), \tag{15}$$

из которого следует, что равенства $F_n = 2$ и $F_p = 2$ должны выполняться при одном напряжении. Дифференцируя (13) и (14) можно получить после несложных преобразований формулы для вычисления $\alpha_p$ при чистой инжекции электронов

$$\alpha_p = \frac{\left[M_n(F_n - 2)\right]'}{2M_n^2(F_n - 1)} \qquad (16)$$

и дырок

$$\alpha_p = \frac{2}{F_p' M_p - (F_p - 2)M_p'}\left(\frac{M_p'}{M_p}\right)^2. \qquad (17)$$

В первом случае $\alpha_n$ можно вычислять по формуле (8), а во втором случае - по формуле (12).

Совместное измерение $M_{n(p)}$ и $F_{n(p)}$ в $n^+ - i - p^+$ – диодах с нелегированным $i$ – слоем также позволяет вычислить $\alpha_{n,p}$. При «чистой» инжекции дырок

$$\alpha_p = \frac{(M_p - 1)^2}{WM_p(M_p - F_p)}\ln\frac{M_p - 1}{F_p - 1}, \qquad \alpha_n = \frac{F_p M_p + 1 - 2M_p}{WM_p(M_p - F_p)}\ln\frac{M_p - 1}{F_p - 1}, \qquad (18)$$

где $W$ - толщина $i$-слоя. При «чистой» инжекции электронов следует заменить индекс «$p$» на «$n$» в (18). Этот метод предложили авторы работы [7], но для расчета $\alpha_{n,p}$ они использовали ошибочные формулы, отличающиеся от (18) аргументом логарифма: в [7] он был равен $(M_p - 1)^2/(M_p - F_p)$. Ошибка возникла из-за использования неправильной формулы для $M_p$, которая на самом деле имеет вид

$$M_p = \frac{\alpha_p - \alpha_n}{\alpha_p \exp\left[(\alpha_n - \alpha_p)w\right] - \alpha_n}.$$

и вместе с формулой $F_p = \alpha_n M_p/\alpha_p + (2 - 1/M_p)(1 - \alpha_n/\alpha_p)$ из [23] приводит к (18). В результате значения $\alpha_{n,p}(E)$, полученные в [7], оказались завышенными, особенно в сильных полях, где отношение $\alpha_n/\alpha_p$ не слишком мало.

Авторы работы [8] использовали для определения $\alpha_{n,p}$ зависимости $M_{an}(U)$ и $M_{dp}(U)$, полученные при «чистой» инжекции электронов в ОПЗ $n^+ - p - p^+$ – диода и дырок в ОПЗ $n^+ - n - p^+$ – диода[2]. Они показали, что в этом случае справедливо соотношение

$$\left[\frac{qN_d}{\varepsilon M_{dp}^2 \alpha_p(E_M)}\frac{dM_{dp}}{dE_M}\right]\left[\frac{qN_a}{\varepsilon M_{an}^2 \alpha_n(E_M)}\frac{dM_{an}}{dE_M}\right]^{N_a/N_d} = 1, \qquad (19)$$

которое позволяет вычислить, например, $\alpha_n$, если известна зависимость $\alpha_p(E_M)$. Эта зависимость, в свою очередь, вычислялась в [8] по приближенной формуле $\alpha_p(E_M) \approx M_p'/M_p$, которую можно использовать, пока $(M_n - 1) \ll 1$. Это, конечно, является недостатком (хотя и незначительным в случае 4H-SiC или Si) метода [8]. Между тем «дополнительные» переходы можно использовать более эффективно. Для этого надо сменить типы проводимостей подложек и освещать фотодиоды со стороны $n^+ - p -$ и $n - p^+$ – переходов, а не со стороны тыловых контактов. В этом случае можно вычислять $\alpha_{n,p}(E_M)$ по формулам

---

[2] Здесь и далее дополнительные индексы «$a$» и «$d$» введены для того, чтобы указать тип примеси, легирующей высокоомный слой диодов, использованных при измерении умножения.

$$\alpha_n(E_M) = \frac{q}{\varepsilon} \frac{(M_{dn}-1)\frac{N_a}{M_{ap}}\frac{dM_{ap}}{dE_M} - \frac{N_d}{M_{dn}}\frac{dM_{dn}}{dE_M}}{M_{dn}M_{ap} - M_{dn} - M_{ap}}, \quad (20)$$

$$\alpha_p(E_M) = \frac{q}{\varepsilon} \frac{(M_{ap}-1)\frac{N_d}{M_{dn}}\frac{dM_{dn}}{dE_M} - \frac{N_a}{M_{ap}}\frac{dM_{ap}}{dE_M}}{M_{dn}M_{ap} - M_{dn} - M_{ap}}, \quad (21)$$

которые следуют из дополняющих друг друга соотношений типа (8). Ясно, что при некотором значении $E_M$ знаменатели в (20) и (21) обращаются в ноль. Однако нетрудно показать, что при любых зависимостях $\alpha_{n,p}(E_M)$ одновременно выполняется условие

$$N_a \frac{dM_{ap}^{-1}}{dE_M} = N_d \frac{dM_{dn}^{-1}}{dE_M}$$

обращения в ноль и числителей в (20) и (21). Это обстоятельство можно использовать для оценки качества измерений коэффициентом умножения.

## Литература